\documentclass[preprint,review,12pt,a4paper,3p]{elsarticle}
\usepackage[version=3]{mhchem} 
\usepackage[]{geometry}
\usepackage[scriptsize,tight]{subfigure}
\usepackage{graphicx}
\usepackage{multirow}
\usepackage{dcolumn}
\usepackage{bm}
\usepackage{color,soul}
\usepackage[mathlines]{lineno}
\usepackage{hyperref}

\biboptions{longnamesfirst,sort&compress,semicolon}

\begin{document}

\begin{frontmatter}

\title{M\"obius carbon nanobelts interacting with heavy metal
nanoclusters}

\author{C. Aguiar$^{1}$}
\author{N. Dattani$^{2,3}$}
\ead{nike@hpqc.org}
\author{I.~Camps\corref{coric}$^{1,3}$}
\ead{icamps@unifal-mg.edu.br}
\cortext[coric]{Corresponding authors}

\address{$^{1}$Laborat\'orio de Modelagem Computacional - \emph{La}Model,
Instituto de Ci\^{e}ncias Exatas - ICEx. Universidade Federal de Alfenas -
UNIFAL-MG, Alfenas, Minas Gerais, Brasil}
\address{$^{2}$HPQC College, Waterloo, Canada}
\address{$^{3}$HPQC Labs, Waterloo, Canada}

\begin{abstract}
To investigate the interaction between carbon and Mobius-type carbon nanobelts
and nickel, cadmium, and lead nanoclusters, we utilized the semiempirical tight
binding framework provided by xTB software. Through our calculations, we
determined the lowest energy geometries, complexes stability, binding energy,
and electronic properties. Our findings demonstrate that heavy metal
nanoclusters have a favorable binding affinity towards both nanobelts, with the
Mobius-type nanobelt having a stronger interaction. Additionally, our
calculations reveal that the nickel nanocluster has the lowest binding energy,
displaying the greatest charge transfer with the nanobelts, which was nearly
twice that of the cadmium and lead nanoclusters.

The molecular dynamic simulation showed that all complexes were stable at
298~K, with low root--mean--square deviation and negative binding energy.
Homogeneous distribution of the frontier orbitals throughout the nanobelt
structure was observed, with slight changes noted in the distribution when the
structure was twisted to create the Mobius-type nanobelt. Furthermore, the
topological study demonstrated that although the number of bonds between the
metal nanoclusters and the Mobius-type nanobelt were the same, the bond
intensities were notably different. Bonds formed with the nickel nanocluster
were stronger than those formed with cadmium and lead metals. Our combined
results lead to the conclusion that the nickel nanoclusters are chemisorbed,
whereas cadmium and lead nanoclusters are physisorbed in both nanobelts.
\end{abstract}


\end{frontmatter}

\section{Introduction}
\label{Sec:Intro}
\newcommand{\sizeA}{3.0cm}

In recent years, major concerns have been raised about pollution caused by
heavy metals resulting from industrialization and daily release into the
environment~\cite{Kumar-Crit.Rev.Env.Sci.Tec.-44-1000-2014,
Hoang-Chemosphere-287-131959-2022}. Improper
disposal of these metals can cause numerous
diseases~\cite{Baby-NanoscaleRes.Lett.-14-341-2019,
Zamora-Ledezma-Environ.Technol.Innov.-22-101504-2021} and harmful effects on
nature~\cite{Kolangare-Environ.Chem.Lett.-17-1053-2018,
Hoang-Chemosphere-287-131959-2022}. Among the heavy metals
commonly found in industrial wastewater are Pb\textsuperscript{2+},
Cd\textsuperscript{2+}, and Ni\textsuperscript{2+}, which have
already demonstrated high levels of
toxicity~\cite{Zwolak-WaterAirSoilPollution-230-1-2019}. Unlike organic
compounds,
heavy metals are not biodegradable, so there is no natural degradation, and
they have been observed in living organisms' tissues and damaged gills of some
fish~\cite{Hoang-J.Mar.Eng.Technol.-20-159-2018,
Schuler-BioScience-68-327-2018,Hoang-Chemosphere-287-131959-2022}.

Therefore, new studies have been developed with the aim of assisting in the
treatment of water with these metals~\cite{Hoang-J.Mar.Eng.Technol.-20-159-2018,
Arora-J.Compos.Sci.-4-135-2020}. Carbon nanotubes (CNTs)
have been
an option and have been cited as adsorbent materials for heavy metals and their
metabolites in contaminated waters~\cite{Arora-J.Compos.Sci.-4-135-2020,
Moallaei-FoodChem.-322-126757-2020,
Ganzoury-J.Hazard.Mater.-393-122432-2020,
Oliveira-Nanomaterials-11-2082-2021,
Hoang-Chemosphere-287-131959-2022}. This is due to the fact
that CNTs
have a large surface area with one hydrophobic side and another easily
modifiable with specific groups~\cite{Hoang-Chemosphere-287-131959-2022,
Murjani-CarbonLetters-32-1207-2022}. Therefore, they can interact in various
ways with heavy metals, such as electrostatic interactions, hydrophobic
effects, and covalent
bonds~\cite{Trevino-Molecules-27-1067-2022,Pyrzynska-Separations-10-152-2023}.

However, the main difficulties for the application of CNTs are their suspension
in treated solutions, pollution caused by chemicals used in functionalization,
and high production costs~\cite{Hoang-Chemosphere-287-131959-2022}. Therefore,
such materials have been reconsidered
in future studies for large-scale application. Due to this, new structures
formed by covalent bonds between benzene rings called cycloparaphenylenes or
carbon nanobelts (CNBs)~\cite{Povie-Science-356-172-2017,
Nishigaki-J.Am.Chem.Soc.-141-14955-2019} have been proposed as alternatives.
These
nanobelts have been investigated as an alternative structure for applications
in various areas, since they have a highly delocalized electronic structure
with \emph{sp\textsuperscript{2}}
hybridization~\cite{Zhang-J.Am.Chem.Soc.-142-1196-2020,
Li-Nanomaterials-13-159-2022}, becoming
attractive
molecules by offering new
possibilities for the synthesis of materials with different functional groups
and serving as a template or seeds for the synthesis of other
structures~\cite{Zhang-J.Am.Chem.Soc.-142-1196-2020,
Seenithurai-Nanomaterials-11-2224-2021,
Li-Nanomaterials-13-159-2022}. In addition, CNBs exhibit properties
that can be exploited when the
belt acquires a M\"obius topology, the M\"obius carbon nanobelt (MCNBs), where
the
``twisted'' structure should exhibit different molecular properties and
movements
compared to that with a normal belt
topology~\cite{Ajami-Nature-426-819-2003,
Segawa-NatureSynthesis-1-535-2022}. Studies show that CBNs
and MCNBs exhibit specific fluorescence and chirality with possible use in
optoelectronic optical materials~\cite{Wu-Angew.Chem.Int.Ed.-60-3994-2020,
Li-Nanomaterials-13-159-2022,
Segawa-NatureSynthesis-1-535-2022,
Li-Angew.Chem.Int.Ed.-61--2022,
} and non--linear applications such
as imaging and optical sensors~\cite{Li-Nanophotonics-7-873-2018,
Li-Nanomaterials-13-159-2022,
Segawa-NatureSynthesis-1-535-2022}.

Thinking about this, properties originating from different topologies can
provide an alternative to solving current problems such as water pollution
caused by chemical contaminants that generate adverse effects on nature and
living organisms~\cite{Baby-NanoscaleRes.Lett.-14-341-2019,
Wu-Environ.Pollut.-246-608-2019}.

In this study, the interaction of Cadmium (Cd), Nickel (Ni), and Lead (Pb),
with carbon and M\"obius-type carbon nanobelts were
investigated using the semiempirical tight binding theory. Several methods were
used to characterize the systems: best interaction
region detection, geometry optimization, molecular dynamics, electronic property
calculations, and topology studies.

\section{Materials and Methods}
\label{Sec:Method}

\begin{figure}[htpb]
\centering
\includegraphics[width=16.5cm]{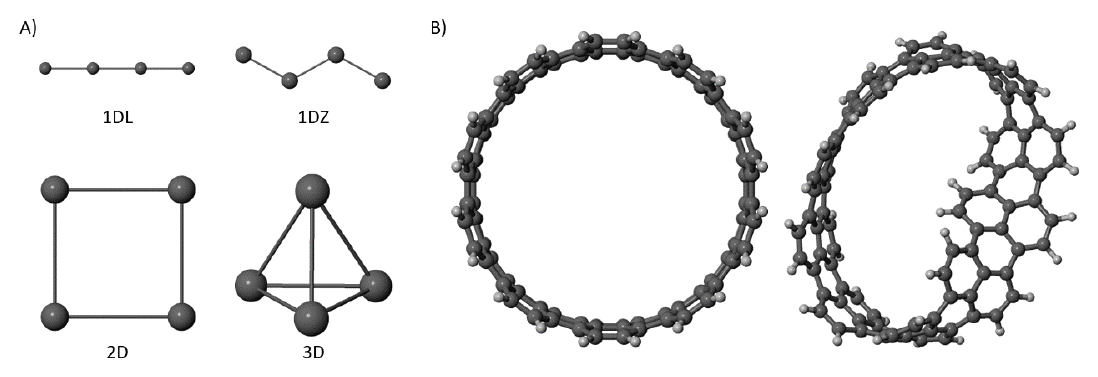}
\caption{Initial structures. Panel A: metal nanoclusters. Panel B: carbon
nanobelt (left),  and M\"obius carbon nanobelt (right).}
\label{Fig:Structures}
\end{figure}

In this work, four different types of four--atom metallic (M4) nanoclusters are
studied interacting with two different types of carbon nanobelts.
Structures of cadmium, nickel and lead clusters were created in the form of a
linear (1DL) and a zigzag (1DZ) one--dimensional chains, a planar
bi--dimensional (2D) structure and a tetrahedron three-dimensional (3D) and were
put to interact with two types of carbon nanostructures:  one consisting on a
nanobelt and the other consisting on a M\"obius nanobelt (twisted nanobelt).
Starting with 2 units of a (10,0) carbon nanosheet repeated 10 times in the z
direction and then wrapped 360 degrees, the nanobelts were generated using the
Virtual NanoLab Atomistix Toolkit software~\cite{VNL}. Then, the periodicity
was removed, and the surface was passivated with hydrogen atoms. For the
M\"obius nanobelts, after the initial repetition of the cells, the nanobelt was
twisted 180 degrees and then wrapped. The structures of the metal nanoclusters,
the carbon nanobelt (CNB) and M\"obius carbon nanobelt (MCNB) are shown in
Figure~\ref{Fig:Structures}.

We utilized the semiempirical tight binding method to carry out the
calculations, which were executed using the xTB program~\cite{xTB_1,xTB_2}. The
steps involved in the calculations are explained in detail below.

Initially, the structures of every single system (comprising two nanobelts and
four nanoclusters for each metal) were optimized. After that, we employed the
automated Interaction Site Screening (aISS)~\cite{xTB-dock} to create various
intermolecular geometries such as CNB+M4 and MCNB+M4. These geometries were
subsequently optimized, where they were ranked using the interaction
energy (xTB-IFF)~\cite{xTB-IFF}. The genetic optimization step was carried out
ten times until the best (lowest interaction energy) complex was obtained. The
top--ranked complexes then underwent structural optimization. To evaluate the
stability of the complexes, each structure from the aISS step underwent a
molecular dynamic (MD) simulation for 100~ps. The parameters used
in geometry optimizations are described in ref.~\cite{myMethods}.

The calculated electronic properties comprised the system energy, the highest
occupied molecular orbital energy (HOMO, $\varepsilon_H$), the lowest
unoccupied molecular orbital energy (LUMO, $\varepsilon_L$), the energy gap
between the HOMO and LUMO orbitals ($\Delta \varepsilon =
\varepsilon_H - \varepsilon_L$), and the atomic charges, which were determined
using the CM5 scheme~\cite{charges_CM5}. Using the computed charges, we
estimated the charge transfer between the nanobelts and the isolated metal
nanocluster by utilizing the following expression

\begin{equation}
\label{Eq:Qtransf}
\Delta {Q_{M4}} = Q_{M4}^{ads} - Q_{M4}^{iso},
\end{equation}
where $Q_{M4}^{ads}$ is the total charge on the metal nanocluster after
adsorption, and $Q_{M4}^{iso}$ is the total charge for the isolated metal
nanocluster.

We computed the binding energies ($E_b$) of the metals adsorbed on the
nanobelts using the subsequent expression:

\begin{equation}
\label{Eq:bind}
E_b = E_{NB+M4} - E_{NB}- E_{M4}.
\end{equation}

In equation~\ref{Eq:bind}, $E_{NB}$ and $E_{M4}$ are the energies
for the isolated nanobelts and metal nanoclusters, respectively, and
$E_{NB+M4}$ is the energy of the NB+M4 complex (CNB+M4 and MCNB+M4 systems).

The MULTIWFN~\cite{multiwfn} software was employed to determine the topological
properties and descriptors (such as critical points, electronic density,
Laplacian of the electronic density, etc.).

\section{Results and discussion}
\label{Sec:results}
\subsection{Nanoclusters adsorption at the BN nanobelts}
\label{Sec:Geometry}

The fully relaxed structures are displayed in Figure~\ref{Fig:OPT_CNB}. Among
both nanobelts, the Cd1DL, Ni2D, and Pb2D nanoclusters exhibited the lowest
energy complexes. Upon initial observation, the interaction with nickel
nanoclusters caused more deformation in the nanobelts compared to the other
metals, and also resulted in more bonds. Both of these tendencies may suggest
that there is stronger binding between CNB and MCNB with Ni.

\renewcommand{\sizeA}{3.0cm}
\begin{figure}[htpb]
\centering
\begin{tabular}{ccc}
\subfigure[CNB+Cd1DL]{\includegraphics[width=\sizeA]{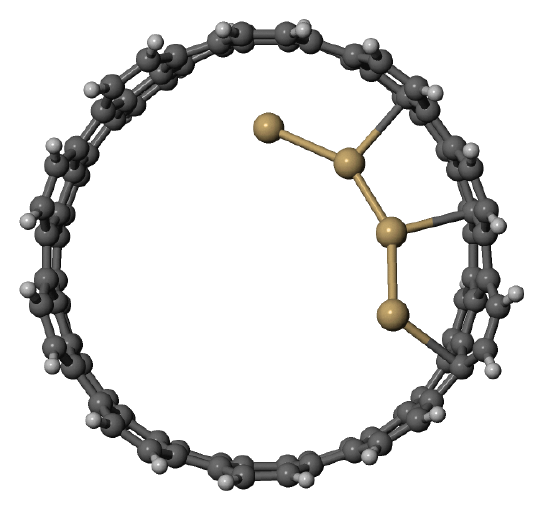}
\label{subFig:CNB+Cd1DL}}
      &
\subfigure[CNB+Ni2D]{\includegraphics[width=\sizeA]{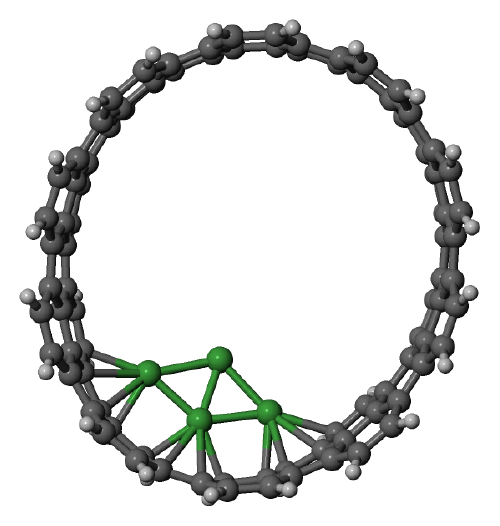}
\label{subFig:CNB+Ni2D}}
      &
\subfigure[CNB+Pb2D]{\includegraphics[width=\sizeA]{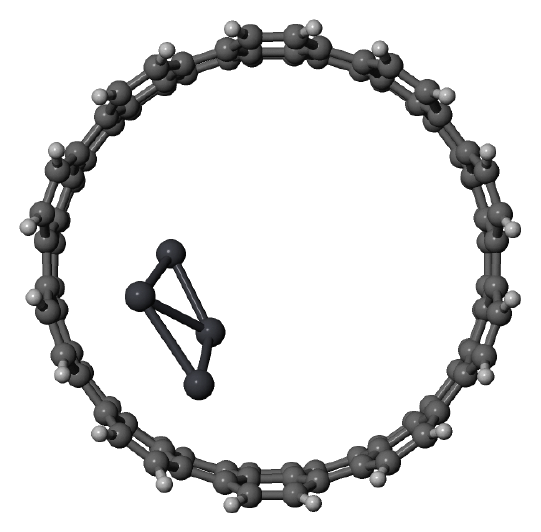}
\label{subFig:CNB+Pb2D}} \\

\subfigure[MCNB+Cd1DL]{\includegraphics[width=\sizeA]{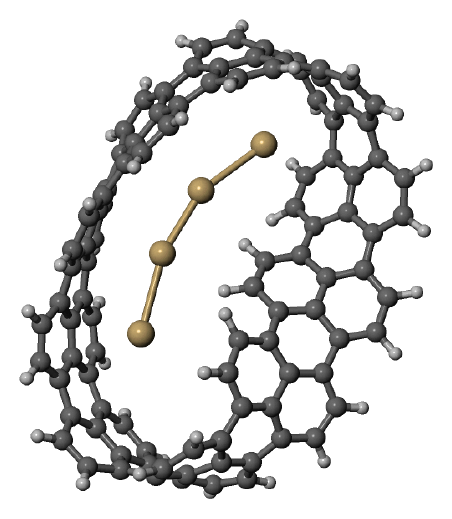}
\label{subFig:MCNB+Cd1DL}}
      &
\subfigure[MCNB+Ni2D]{\includegraphics[width=\sizeA]{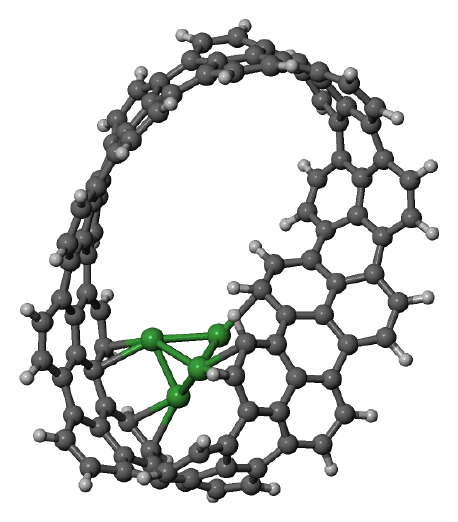}
\label{subFig:MCNB+Ni2D}}
      &
\subfigure[MCNB+Pb2D]{\includegraphics[width=\sizeA]{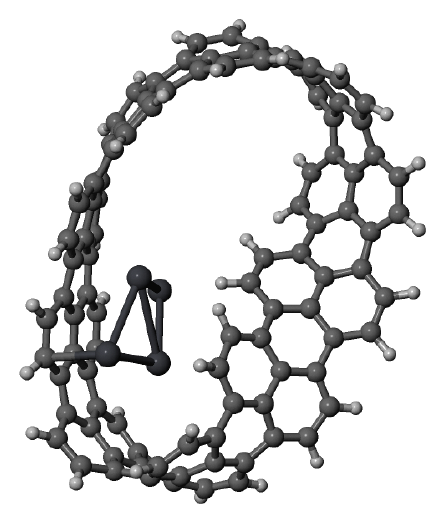}
\label{subFig:MCNB+Pb2D}}

\end{tabular}
\caption{\label{Fig:OPT_CNB}  Fully relaxed complexes with lowest energies.
Image rendered with Jmol software~\cite{Jmol}.}
\end{figure}

One of the primary differences between chemisorption and physisorption is their
effect on the electronic states of both the adsorbate and the
adsorbent~\cite{Everett2001}. Sections~\ref{Sec:ElecProp}, and~\ref{Sec:Topo}
will show that the electronic properties of the systems are modified after
adsorption, which confirms this phenomenon.

\begin{table}[htpb]
\caption{Electronic and bond distances from optimized geometries$^\dagger$.}
\label{Tab:DataResults}
\begin{center}
\begin{tabular}{lrrrrrr}
  \hline
  Complex & $E_{b}^{aISS}$ & $E_b$ & $\Delta Q_M$ & Gap ($\Delta \varepsilon$)
  & Distances \\
  \hline
  \hline
  CNB+Cd1DL      &  -77.29   &  -59.38   & 0.1517 & 0.632 &
  2.79/2.80/2.81/3.06/3.10\\

  CNB+Ni2D    &  -136.25   &  -123.79  &  -0.7671  & 0.158  &
  2.36(2)/2.37(2)/2.38(2)/2.43(2)\\
  CNB+Pb2D    &  -100.83   & -61.78   &  0.3862  & 0.039  & 2.84/2.86/3.74\\
  \hline
  MCNB+Cd1DL     &  -91.74   & -77.91   & 0.2328   & 0.543  &
  2.80/3.00/3.01/3.14\\
                &           &          &          &         &
                 3.23/3.27/3.36/3.41/3.45\\
  MCNB+Ni2D    & -158.77    & -146.30   & -0.7426   & 0.101  &
  2.35/2.37/2.42/2.44/2.45\\
  MCNB+Pb2D    &  -118.60   & -79.56   & 0.8055   & 0.495  &
  2.41/2.67/2.68/2.76/2.80/3.32/3.58\\
  \hline
\end{tabular}
\begin{flushleft}
\tiny {$^\dagger$ $E_{b}^{aISS}$ and $E_{b}$ are in units kcal/mol, $\Delta Q_M$
are in units of $e$, $\Delta \varepsilon$ is in units of $eV$ and the distances
are in units of \AA, respectively.}
\end{flushleft}
\end{center}
\end{table}

Table~\ref{Tab:DataResults} presents the data for the complexes with the lowest
energy. The table reports two binding energies: $E_{b}^{aISS}$ and $E_b$. The
former is obtained from the aISS step, while the latter is calculated using
equation~\ref{Eq:bind} for the fully optimized individual structures and the
final complex. Upon comparing both binding energies for each system, it is
observed that the Ni2D nanocluster has the lowest binding energies for both
nanobelts, which suggests stronger adsorption. The binding energy ($E_b$) for
Cd1DL and Pb2D clusters are very similar with a difference around 2~kcal/mol.
Since all the binding energies are negative, it can be inferred that all the
adsorption processes are favorable.

The visual depiction of the complexes in Figure~\ref{Fig:OPT_CNB} suggests that
only Cd and Ni
established bonds with the nanobelts. Specifically, Cd formed three bonds with
CNB and none with MCNB, while the Ni cluster formed twelve bonds with CNB and
six with MCNB. In contrast, Pb did not establish any bonds with either CNB or
MCNB. However, since the bond information from Figure~\ref{Fig:OPT_CNB} is
solely based on
geometrical data, the Quantum Theory of Atoms in Molecule
(QTAIM)~\cite{bader1994} was
employed in Section~\ref{Sec:Topo} as a more precise method to examine bond
formation.

Geometry optimization entails utilizing an algorithm to acquire a local minimum
structure on the potential energy surface (PES) to determine the lowest energy
conformers of a system. However, this approach does not offer any insights into
the system's stability over time. In contrast, molecular dynamics simulation
examines the motion of atoms and molecules at a specific temperature (in this
case, 298.15~K) and provides a way to explore the PES. We utilized this method
to perform simulations on each complex, starting with the structures obtained
from the aISS step as initial conformations. The simulations were conducted for
a production time frame of 100~ps with a time step of 2~fs and an optional dump
step of 50~fs, at which the final structure was saved to a trajectory file.

In Figures~\ref{Fig:MD_CNB} and~\ref{Fig:MD_MCNB}, panel A displays the
root--mean--square deviation (RMSD) of the metal nanoclusters, while panel B
shows the system frames at various simulation times (0~ps, 25~ps, 50~ps, 75~ps,
and 100~ps). The low RMSD values ($<$ 2~\AA) suggest that the metal nanocluster
conformations remained relatively constant, indicating structural stability.
By comparing the figures' snapshots with
Figure~\ref{Fig:Structures}B, we confirmed that the Ni nanocluster caused the
most significant changes to the nanobelt. From panel B, it is apparent that the
metal nanoclusters remained bound to their corresponding carbon nanobelts in
all cases, indicating complex stability. This is confirmed by the negative
binding energy during the simulation time as shown in
Figure~\ref{Fig:BindingEnergy}. The full molecular dynamics movies
can be downloaded from the Zenodo server~\cite{aguiar_c_2023_7821847}.

\renewcommand{\sizeA}{14.0cm}
\begin{figure}[htpb]
\centering
\includegraphics[width=\sizeA]{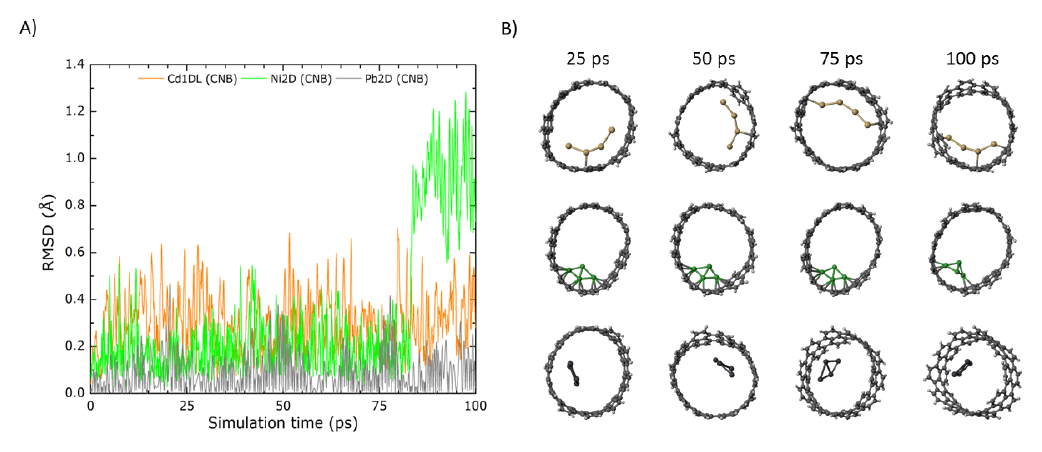}
\label{Fig:RMSD_CNB}
\caption{\label{Fig:MD_CNB} Panel A: Calculated RMSD for metal nanocluster from
CNB complexes.  Panel B: Snapshots of CNB complexes at different simulation
times.}
\end{figure}

\renewcommand{\sizeA}{14.0cm}
\begin{figure}[htpb]
\centering
\includegraphics[width=\sizeA]{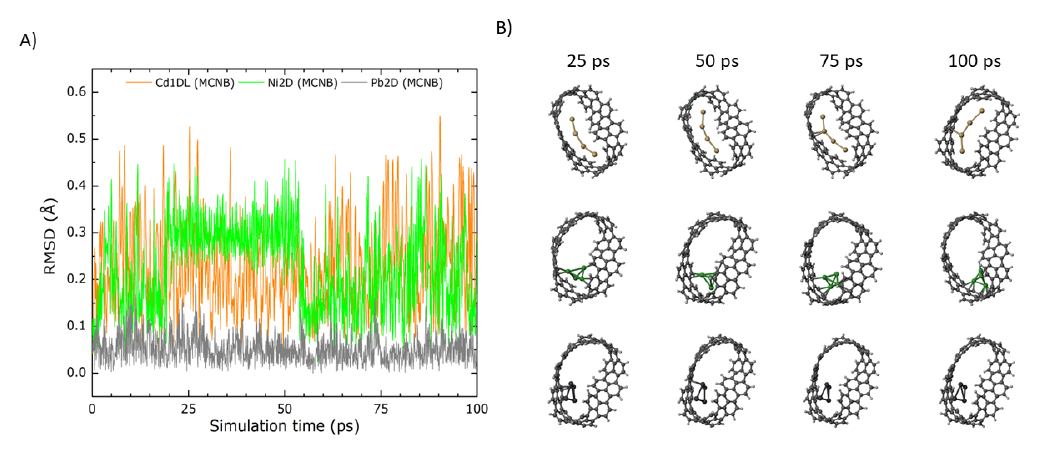}
\label{Fig:RMSD_MCNB}
\caption{\label{Fig:MD_MCNB} Panel A: Calculated RMSD for metal nanocluster
from MCNB complexes.  Panel B: Snapshots of MCNB complexes at different
simulation times.}
\end{figure}

\renewcommand{\sizeA}{7.0cm}
\begin{figure}[htpb]
\centering
\subfigure[CNB]{\includegraphics[width=\sizeA]{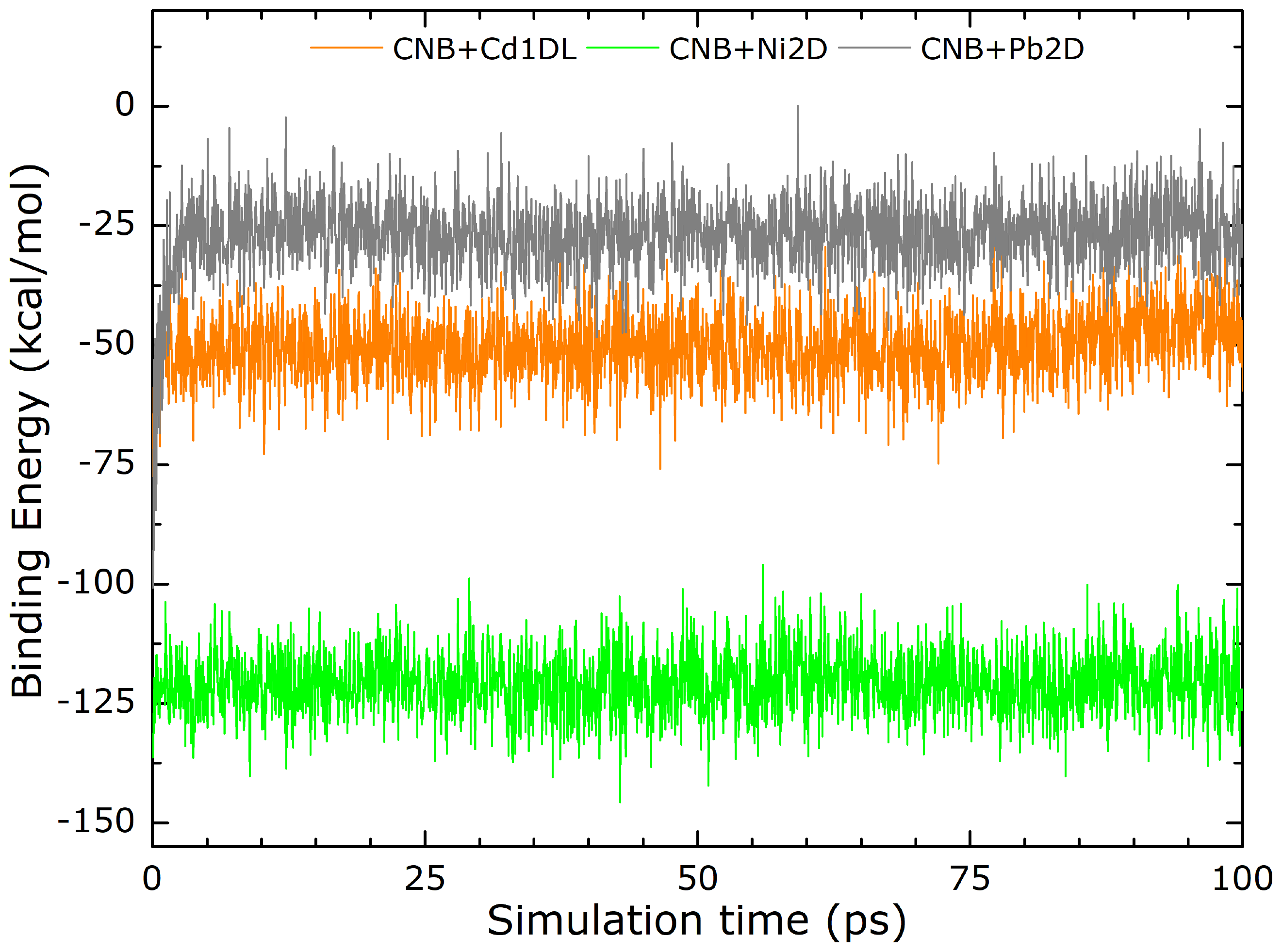}
\label{Fig:Eb_CNB}}
\subfigure[MCNB]{\includegraphics[width=\sizeA]{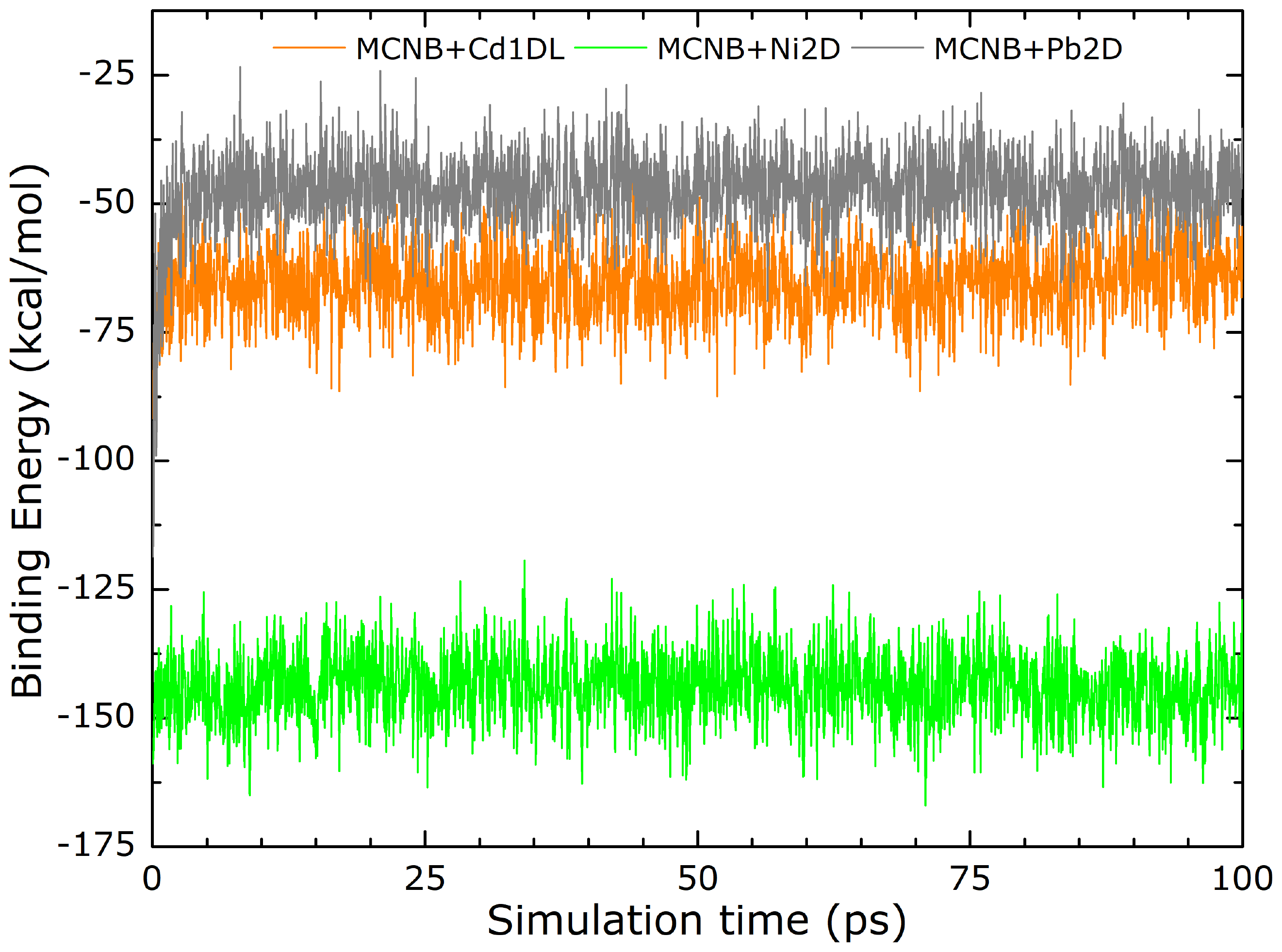}
\label{Fig:Eb_MCNB}}
\caption{\label{Fig:BindingEnergy} Binding energy of CNB and MCNB complexes
during the simulation time.}
\end{figure}

\subsection{Electronic properties}
\label{Sec:ElecProp}

Figures~\ref{Fig:HOMO} and~\ref{Fig:LUMO} show the calculated highest
occupied/lowest unoccupied molecular orbitals (HOMO/LUMO) for all optimized
systems.

The molecular orbitals for the CNB system (Figures~\ref{Fig:HOMO_CNB}
and~\ref{Fig:LUMO_CNB}) are homogeneously distributed over the entire
structure, as expected due to the belt symmetry. Surprisingly, despite the
symmetry break due to the twist, the HOMO and LUMO distribution for MCNB
(Figures~\ref{Fig:HOMO_MCNB} and~\ref{Fig:LUMO_MCNB}) remain
almost homogeneously distributed over the M\"obius carbon nanobelt. This
behavior is in contrast what happens for M\"obius boron--nitride nanobelt where
the HOMO/LUMO distributions are affected by the twist~\cite{aguiar-arxiv-MBNNB}.

Figures~\ref{Fig:HOMO} and~\ref{Fig:LUMO} depict the HOMO/LUMO molecular
orbitals for all optimized systems. The CNB system's molecular orbitals
(Figures~\ref{Fig:HOMO_CNB} and~\ref{Fig:LUMO_CNB}) are distributed uniformly
throughout the structure, as expected due to the symmetry of the belt.
Surprisingly, despite the symmetry break caused by the twist, the HOMO and LUMO
distribution for MCNB (Figures~\ref{Fig:HOMO_MCNB} and~\ref{Fig:LUMO_MCNB})
remain almost uniformly distributed over the M\"obius carbon nanobelt. The
finding presented here is in contrast to the behavior of M\"obius
boron--nitride nanobelts, where the twist affects the HOMO/LUMO distributions,
as reported in our previous study~\cite{aguiar-arxiv-MBNNB}. This indicates
that the electron distribution in carbon nanobelts is more flexible or
``plastic'' compared to boron--nitride nanobelts. The electronic gap for
the CNB is equal to 0.449~eV, which is slightly decreased to 0.352~eV for the
MCNB.

When metal nanoclusters bind to carbon nanobelts, it causes a slight decrease
in the volume of the frontier orbitals on the nanobelt's surface, with the most
significant decrease observed for Ni, followed by Cd and then Pb. This is
unlike what occurs when the same metal nanoclusters interact with boron-nitride
nanobelts, where the binding drastically changes the orbital surface
distribution~\cite{aguiar-arxiv-MBNNB}. Additionally, the interaction between
the nanobelts and the metals increases the volume of the orbitals surfaces
around the regions where the metals are bonded. The interaction of both
nanobelts with all the metals modified the gap, as shown in
Table~\ref{Tab:DataResults}. The greatest charge transfer ($\Delta
{Q_{M4}}$) between the nanobelts and the nanoclusters occurred in the case of
Ni for both CNB and MCNB. Based on the significant modifications observed in
the electronic properties of the CNB and MCNB after interaction with metals, it
can be concluded that chemisorption occurs for all systems.

\renewcommand{\sizeA}{4.0cm}
\begin{figure}[tbph]
\centering
\begin{tabular}{cccc}
\subfigure[CNB]{\includegraphics[width=\sizeA]{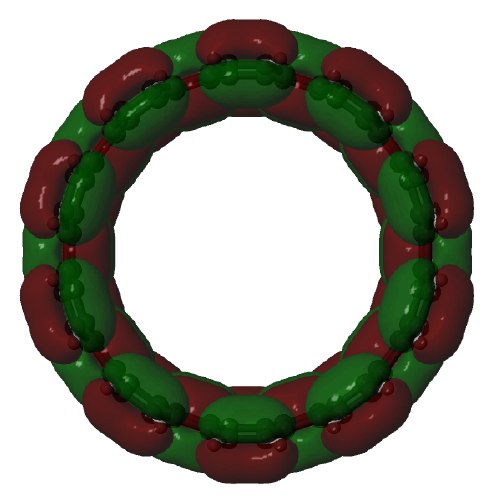}
\label{Fig:HOMO_CNB}}      &
\subfigure[CNB+Cd1DL]{\includegraphics[width=\sizeA]{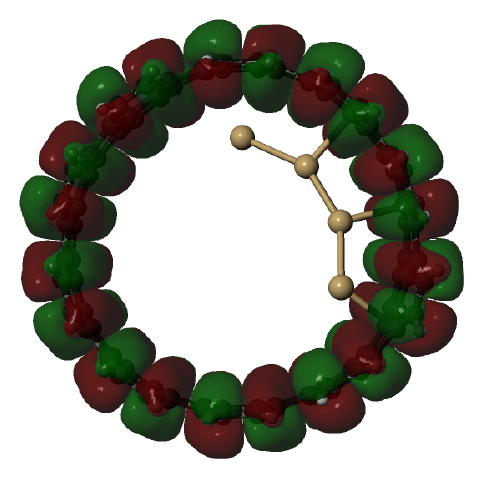}
\label{Fig:HOMO_CNB+Cd1DL}}      &
\subfigure[CNB+Ni2D]{\includegraphics[width=\sizeA]{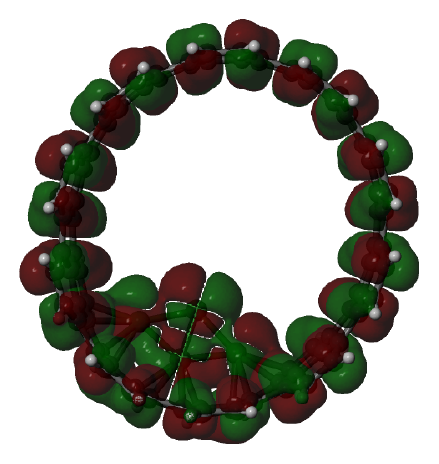}
\label{Fig:HOMO_CNB+Ni2D}}      &
\subfigure[CNB+Pb2D]{\includegraphics[width=\sizeA]{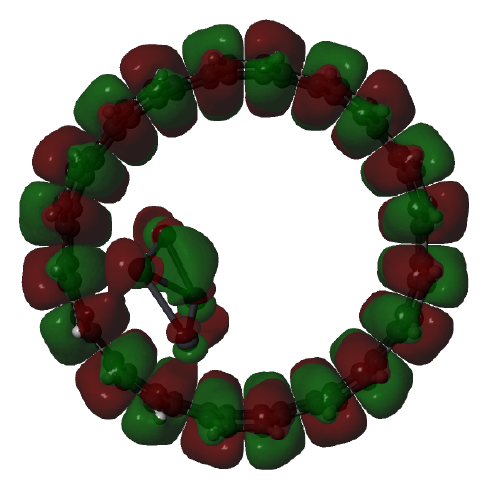}
\label{Fig:HOMO_CNB+Pb2D}} \\

\subfigure[MCNB]{\includegraphics[width=\sizeA]{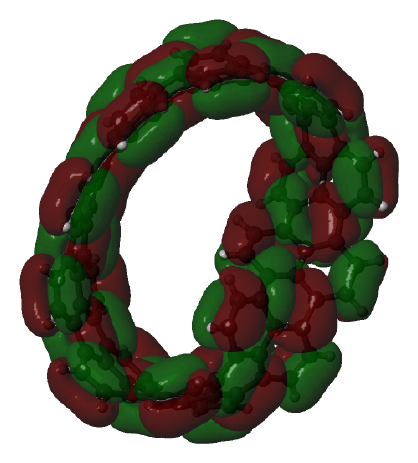}
\label{Fig:HOMO_MCNB}}      &
\subfigure[MCNB+Cd1DL]{\includegraphics[width=\sizeA]{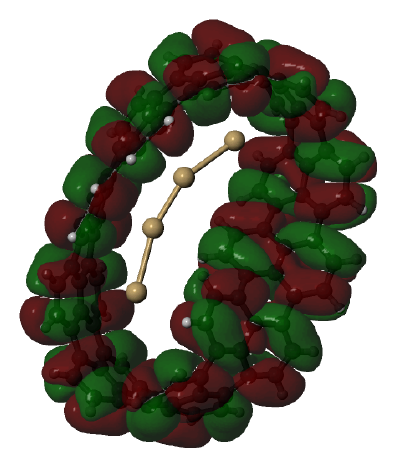}

\label{Fig:HOMO_MCNB+Cd1DL}}      &
\subfigure[MCNB+Ni2D]{\includegraphics[width=\sizeA]{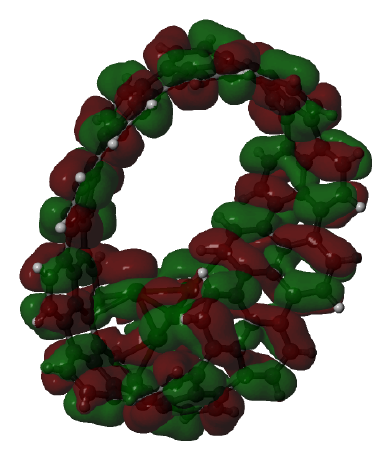}
\label{Fig:HOMO_MCNB+Ni2D}}      &
\subfigure[MCNB+Pb2D]{\includegraphics[width=\sizeA]{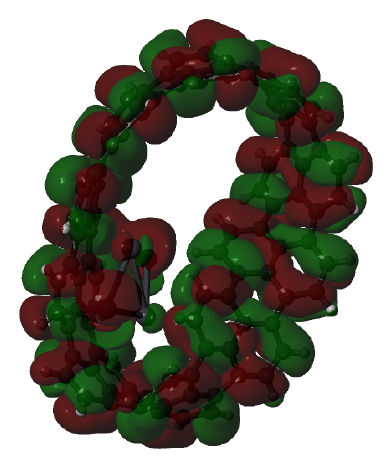}
\label{Fig:HOMO_MCNB+Pb2D}} \\
\end{tabular}
\caption{\label{Fig:HOMO} Highest occupied molecular orbital (HOMO) for all
systems. Red (green) color
represents negative (positive) values. Orbital surfaces rendered with with
isovalue equal to 0.001 and with Jmol software~\cite{Jmol}.}
\end{figure}

\renewcommand{\sizeA}{4.0cm}
\begin{figure}[tbph]
\centering
\begin{tabular}{cccc}
\subfigure[CNB]{\includegraphics[width=\sizeA]{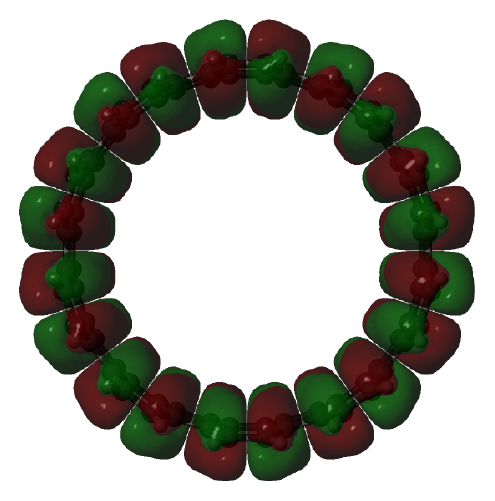}
\label{Fig:LUMO_CNB}}      &
\subfigure[CNB+Cd1DL]{\includegraphics[width=\sizeA]{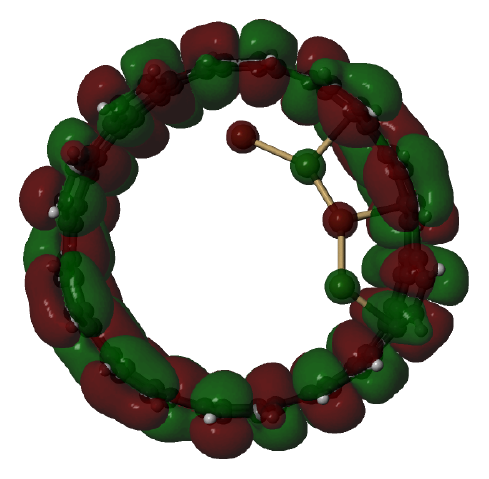}
\label{Fig:LUMO_CNB+Cd1DL}}      &
\subfigure[CNB+Ni2D]{\includegraphics[width=\sizeA]{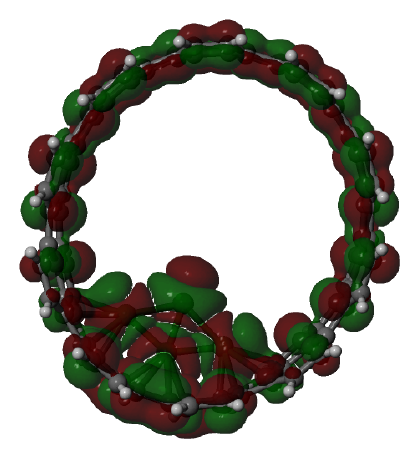}
\label{Fig:LUMO_CNB+Ni2D}}      &
\subfigure[CNB+Pb2D]{\includegraphics[width=\sizeA]{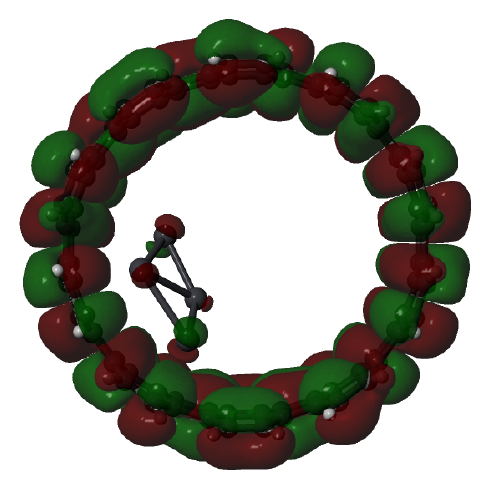}
\label{Fig:LUMO_CNB+Pb2D}} \\
\subfigure[MCNB]{\includegraphics[width=\sizeA]{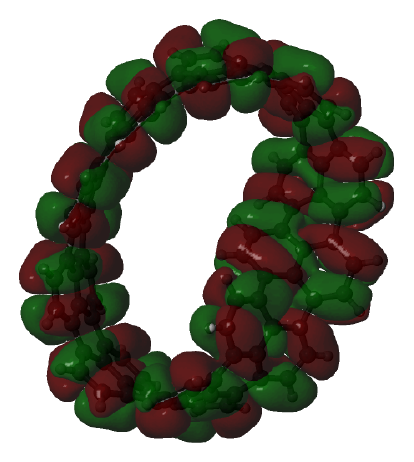}
\label{Fig:LUMO_MCNB}}      &
\subfigure[MCNB+Cd1DL]{\includegraphics[width=\sizeA]{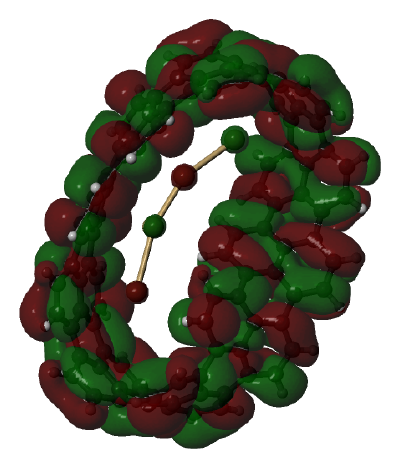}
\label{Fig:LUMO_MCNB+Cd1DL}}      &
\subfigure[MCNB+Ni2D]{\includegraphics[width=\sizeA]{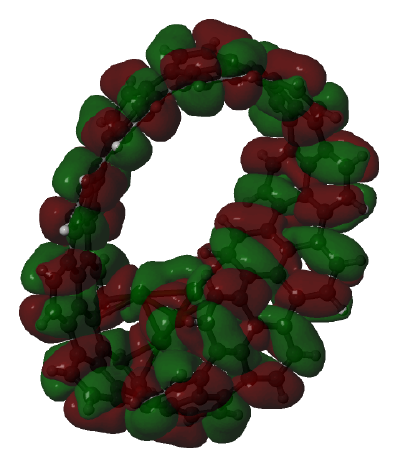}
\label{Fig:LUMO_MCNB+Ni2D}}      &
\subfigure[MCNB+Pb2D]{\includegraphics[width=\sizeA]{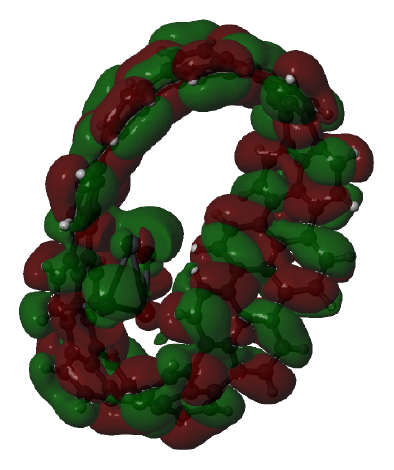}
\label{Fig:LUMO_MCNB+Pb2D}} \\
\end{tabular}
\caption{\label{Fig:LUMO} Lowest unoccupied molecular orbital (LUMO) for all
systems. Red (green) color
represents negative (positive) values. Orbital surfaces rendered with with
isovalue equal to 0.001 and with Jmol software~\cite{Jmol}.}
\end{figure}

\subsection{Topological analysis}
\label{Sec:Topo}

The aim of topological analysis is to detect critical points, which are
positions where the gradient norm of the electron density value equals zero.
The critical points are categorized into four types based on the negative
eigenvalues of the Hessian matrix of the real function~\cite{bader1994}. The
criteria for bond classification and types of critical points are described in
detail in ref.~\cite{myMethods}.

Figure~\ref{Fig:3D-CPs} show the critical points for each complex. The orange
dots represent the bond critical points
(BCPs), the yellow dots represent the ring critical points (RCPs), and the
green dots represent the cage critical points (CCPs). All the complexes show
several
critical points, indicating a favorable interaction between the metals and the
belts. In all cases, the number of critical points made between the metal
nanocluster and the M\"obious nanobelts is greater than with the nanobelts
alone. This can be associated with the fact that M\"obious belts form like small
pockets where the metal nanocluster can be docked. The interaction between the
metal nanocluster with the MCNB create bonds with both sides of the pocket. For
better visualization of the formed critical points, movies with spinning
structures can be downloaded from the Zenodo
server~\cite{aguiar_c_2023_7821847}. Another confirmation
that the strongest interactions are between the Ni nanoclusters and the
carbon nanobelts is the lowest bond distances shown in
Table~\ref{Tab:DataResults}.

Figure~\ref{Fig:3D-CPs} displays the critical points for each complex, with the
bond critical points (BCPs) represented by orange dots, the ring critical
points (RCPs) represented by yellow dots, and the cage critical points (CCPs)
represented by green dots. The presence of multiple critical points in all
complexes indicates a favorable interaction between the metals and the belts.
Moreover, the number of critical points formed between the metal nanocluster
and the Möbious nanobelts is higher than those formed between the nanobelts
alone, which can be attributed to the small pockets formed by the Möbious belts
that allow for the metal nanocluster to dock. The interaction between the metal
nanocluster and the MCNB leads to bonds formed on both sides of the pocket. For
better visualization of the formed critical points, spinning structure movies
can be downloaded from the Zenodo server~\cite{aguiar_c_2023_7821847}. The
lowest bond distances shown
in Table~\ref{Tab:DataResults} provide additional evidence that the strongest
interactions occur
between the Ni nanoclusters and the carbon nanobelts.

\renewcommand{\sizeA}{4.0cm}
\begin{figure}[tbph]
\centering
\begin{tabular}{cccc}
\subfigure[CNB+Cd1DL]{\includegraphics[width=\sizeA]{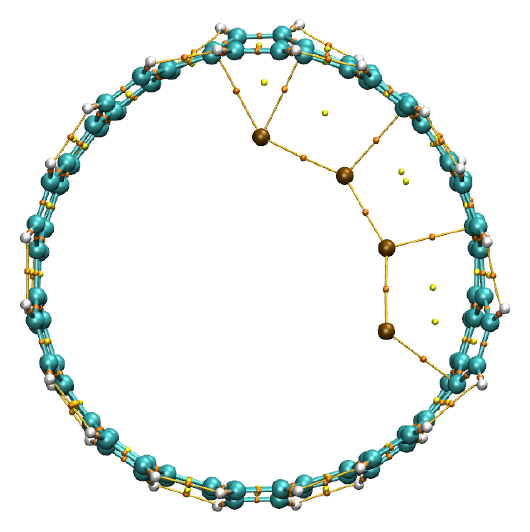}
\label{Fig:topo_CNB+Cd1DL}}      &
\subfigure[CNB+Ni2D]{\includegraphics[width=\sizeA]{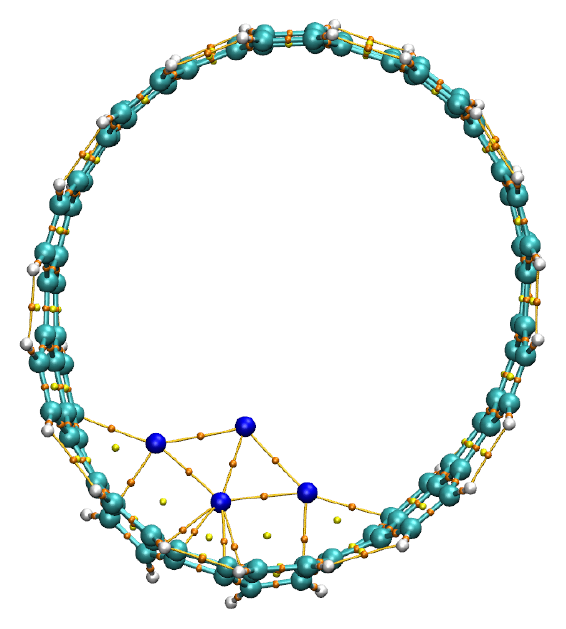}
\label{Fig:topo_CNB+Ni2D}}      &
\subfigure[CNB+Pb2D]{\includegraphics[width=\sizeA]{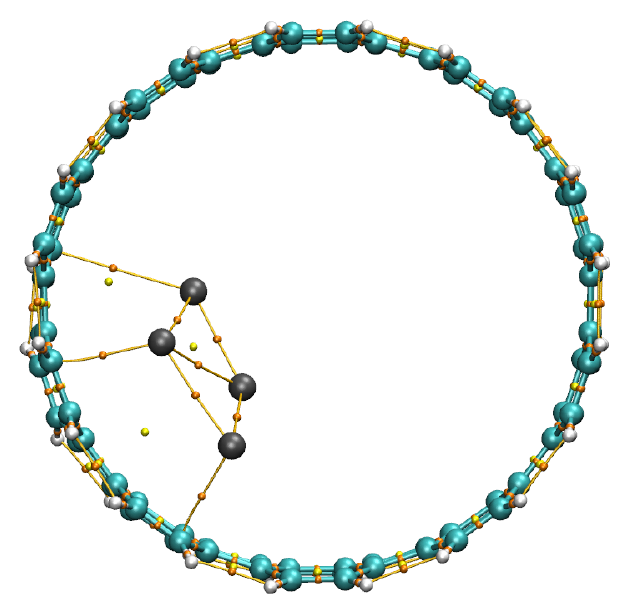}
\label{Fig:topo_CNB+Pb2D}} \\
\subfigure[MCNB+Cd1DL]{\includegraphics[width=\sizeA]{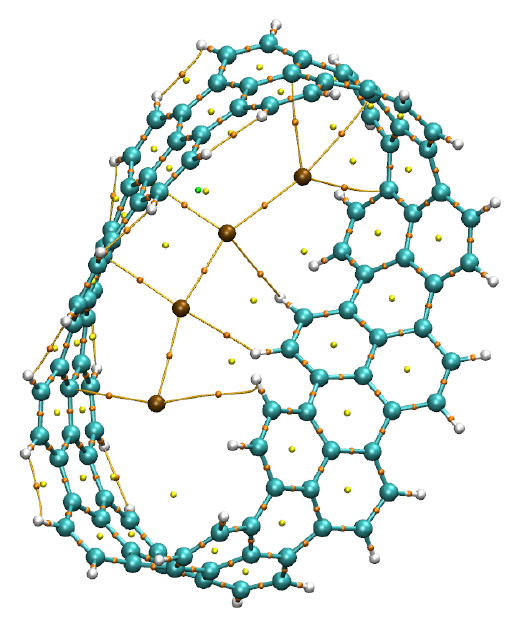}
\label{Fig:topo_MCNB+Cd1DL}}      &
\subfigure[MCNB+Ni2D]{\includegraphics[width=\sizeA]{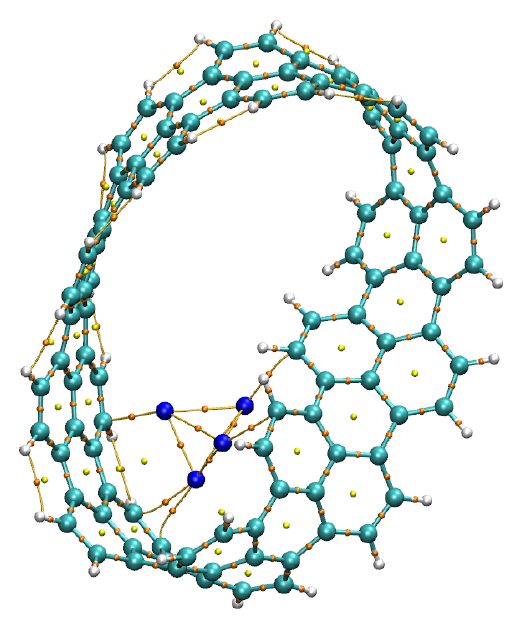}
\label{Fig:topo_MCNB+Ni2D}}      &
\subfigure[MCNB+Pb2D]{\includegraphics[width=\sizeA]{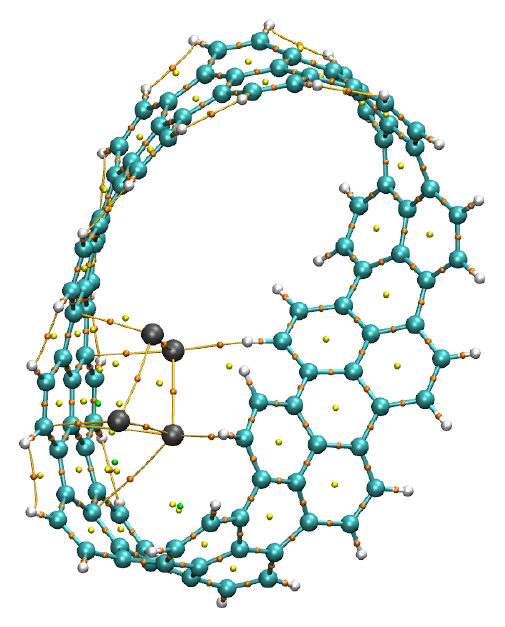}
\label{Fig:topo_MCNB+Pb2D}} \\
\end{tabular}
\caption{\label{Fig:3D-CPs} Critical points: BCPs, RCPs and CCPs (orange,
yellow, and green dots, respectively). Image rendered with VMD
software~\cite{vmd}.}
\end{figure}

The use of $\rho$ and $\nabla^2\rho$ values and indexes such as ELF and LOL can
provide insight into the bond type (covalent or non-covalent) in various
systems. The localization of electron movement is related to the ELF index,
which ranges from 0 to 1~\cite{elf,elf2}. High values of the ELF index indicate
a high degree of electron localization, which suggests the presence of a
covalent bond. The LOL index is another function that can be used to identify
regions of high localization~\cite{lol}. The LOL index also ranges from 0 to 1,
with smaller (larger) values usually occurring in the boundary (inner) regions.

Figure~\ref{Fig:Topol} shows the electron density ($\rho$), Laplacian of the
electron density ($\nabla ^2\rho$), electron localization function (ELF) index,
and localized orbital locator (LOL) index at all the detected bond critical
points. The values of BCPs descriptors for MCNB is greater than for CNB,
indicating that using M\"obius carbon nanobelts to capture heavy metal
nanoclusters is a better choice.

As higher values of $\rho$ can be an indicator of the strength of the bond,
Figures~\ref{Fig:Rho_CNB},~\ref{Fig:Lap_CNB},~\ref{Fig:Rho_MCNB},
and~\ref{Fig:Lap_MCNB} show that MCNB made stronger bonds with Ni than with Cd
or Pb (green, orange and gray regions, respectively). Even though Cd
nanocluster forms more bonds with MCNB than Ni nanocluster,
they have, in general, lower values of $\rho$ and $\nabla^2\rho$ except for one
bond. The ELF (figures~\ref{Fig:ELF_CNB}, and~\ref{Fig:ELF_MCNB}) and LOL
(figures~\ref{Fig:LOL_CNB}, and~\ref{Fig:LOL_MCNB}) indexes show that only one
Cd bond critical point has greater values than for Ni and Pb clusters.
The topological analysis confirms that both carbon nanobelts are capable of
adsorbing the three heavy metal nanoclusters studied here. Nevertheless, the
MCNB presented greater values for all the descriptors used. In all cases, the
Ni nanoclusters are chemisorbed, whereas Cd and Pb nanoclusters are physisorbed.

Figure~\ref{Fig:Topol} displays several indexes related to the electron density
and its confinement, such as the electron localization function (ELF) and
localized orbital locator (LOL) index, for all the identified bond critical
points. The descriptors for bond critical points (BCPs) are greater for
M\"obius carbon nanobelts (MCNB) compared to conventional carbon nanobelts
(CNB), indicating that MCNB is more suitable for capturing heavy metal
nanoclusters. The figures of the electron density ($\rho$) and Laplacian of the
electron density ($\nabla ^2\rho$),
Figures~\ref{Fig:Rho_CNB},\ref{Fig:Lap_CNB},\ref{Fig:Rho_MCNB},
and~\ref{Fig:Lap_MCNB}, show that MCNB makes stronger bonds with Ni than with
Cd or Pb, as shown by the larger bars of green, orange, and gray colors,
respectively. While Cd nanocluster forms more bonds with MCNB than Ni
nanocluster, their values of $\rho$ and $\nabla^2\rho$ are generally lower,
except for one bond. The ELF and LOL indexes for
Figures~\ref{Fig:ELF_CNB},\ref{Fig:ELF_MCNB},\ref{Fig:LOL_CNB},
and~\ref{Fig:LOL_MCNB} reveal that only one Cd bond critical point has greater
values than the critical points descriptors for Ni and Pb clusters. The
analysis indicates that both types of
carbon nanobelts can adsorb the three heavy metal nanoclusters studied here.
Nonetheless, MCNB shows greater values for all the descriptors used. In all
cases, Ni nanoclusters are chemisorbed, whereas Cd and Pb nanoclusters are
physisorbed.

\renewcommand{\sizeA}{3.5cm}
\begin{figure}[tbph]
\centering
\begin{tabular}{ccccc}
\subfigure[CNB+M4 ($\rho$)]{\includegraphics[width=\sizeA]{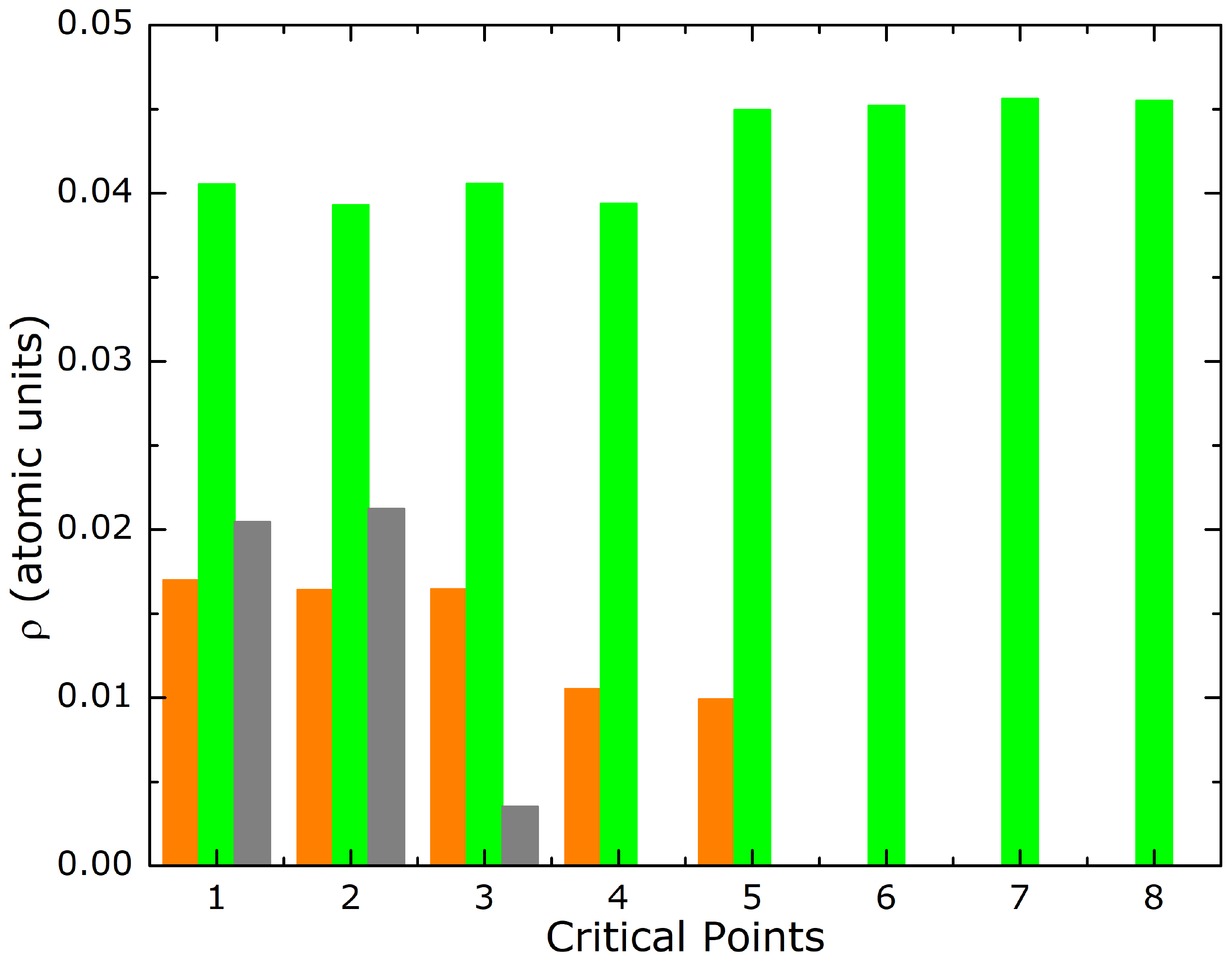}
\label{Fig:Rho_CNB}}      &
\subfigure[CNB+M4 ($\nabla
^2\rho$)]{\includegraphics[width=\sizeA]{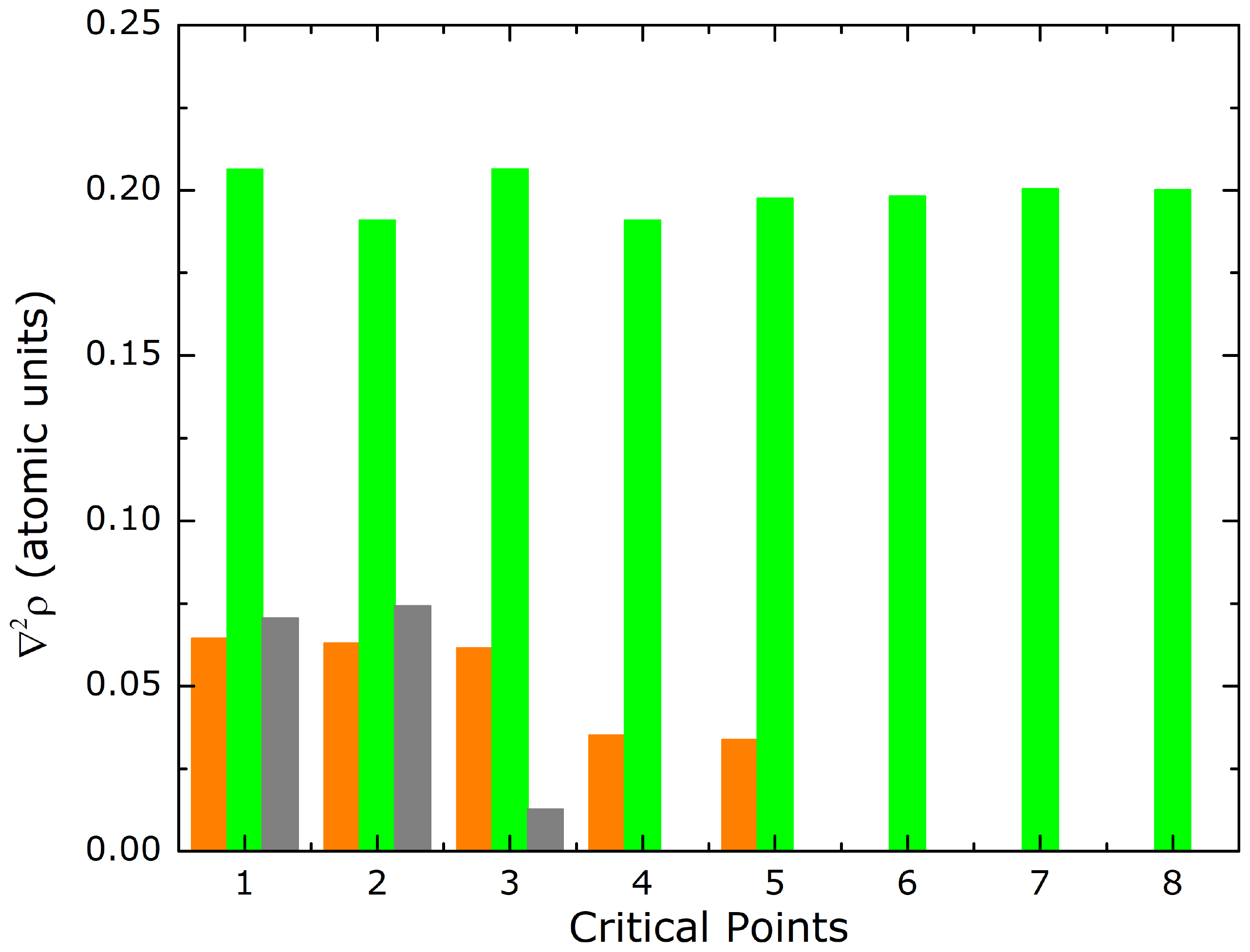}
\label{Fig:Lap_CNB}}      &
\subfigure[CNB+M4 (ELF)]{\includegraphics[width=\sizeA]{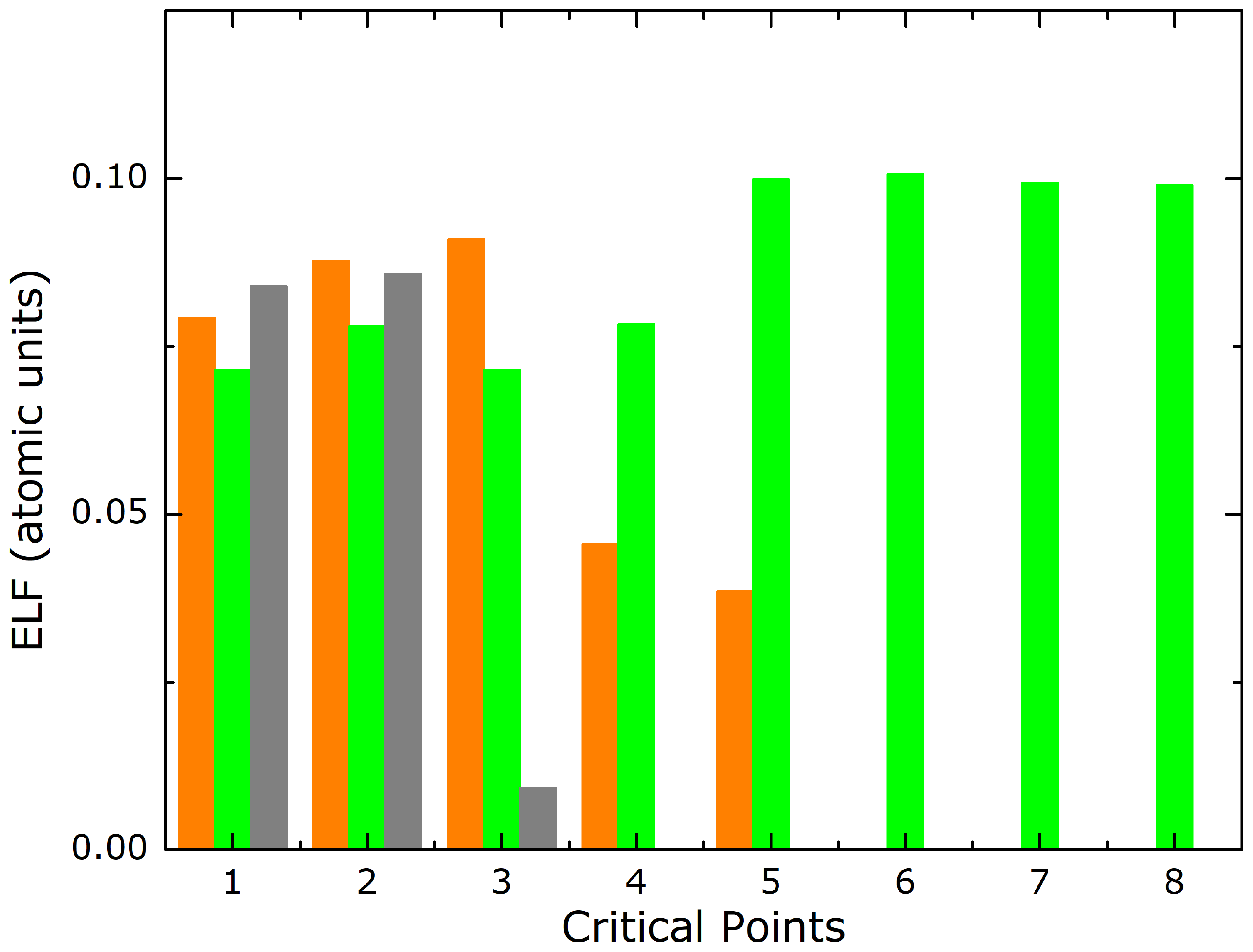}
\label{Fig:ELF_CNB}}       &
\subfigure[CNB+M4 (LOL)]{\includegraphics[width=\sizeA]{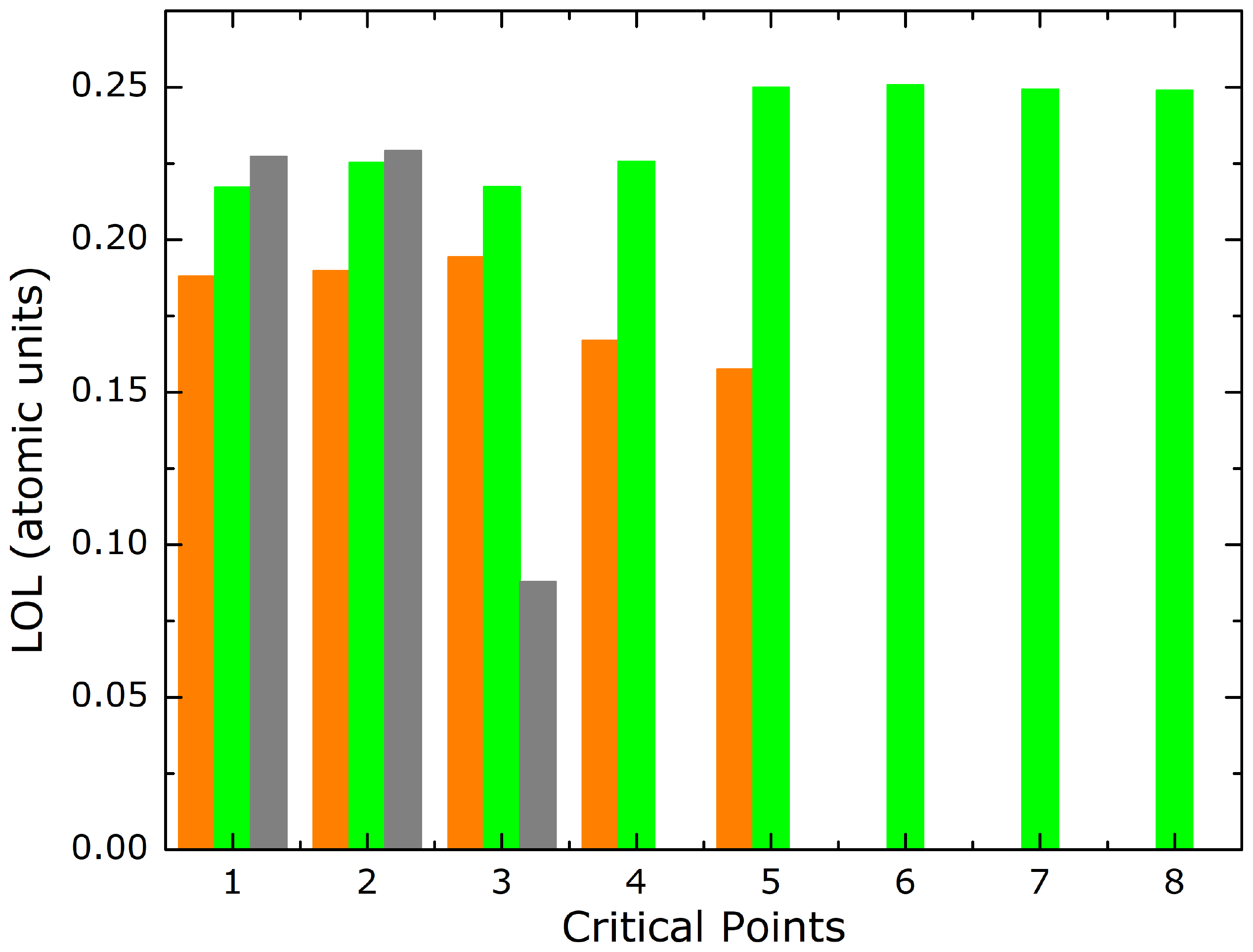}
\label{Fig:LOL_CNB}}\\
\subfigure[MCNB+M4 ($\rho$)]{\includegraphics[width=\sizeA]{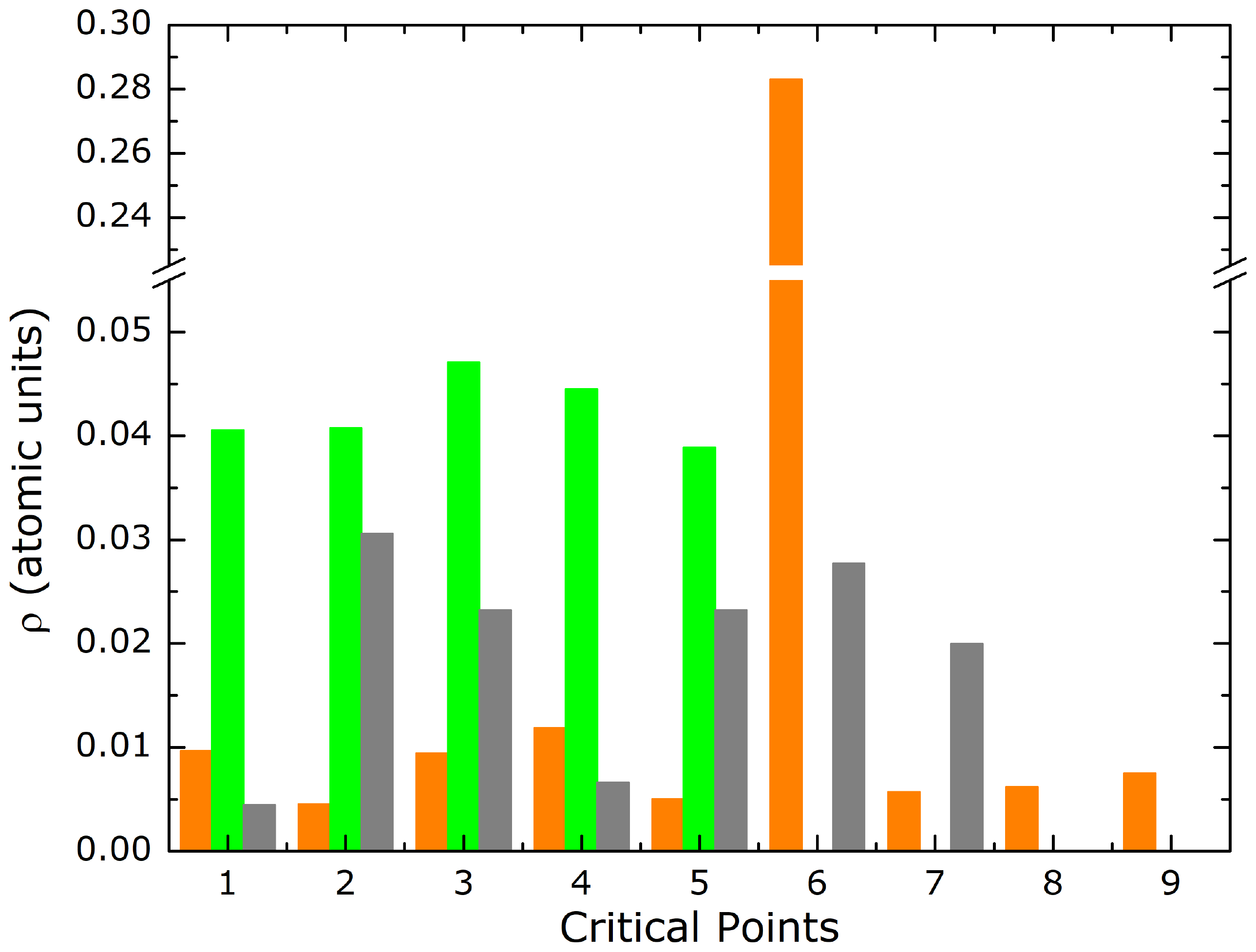}
\label{Fig:Rho_MCNB}}      &
\subfigure[MCNB+M4 ($\nabla
^2\rho$)]{\includegraphics[width=\sizeA]{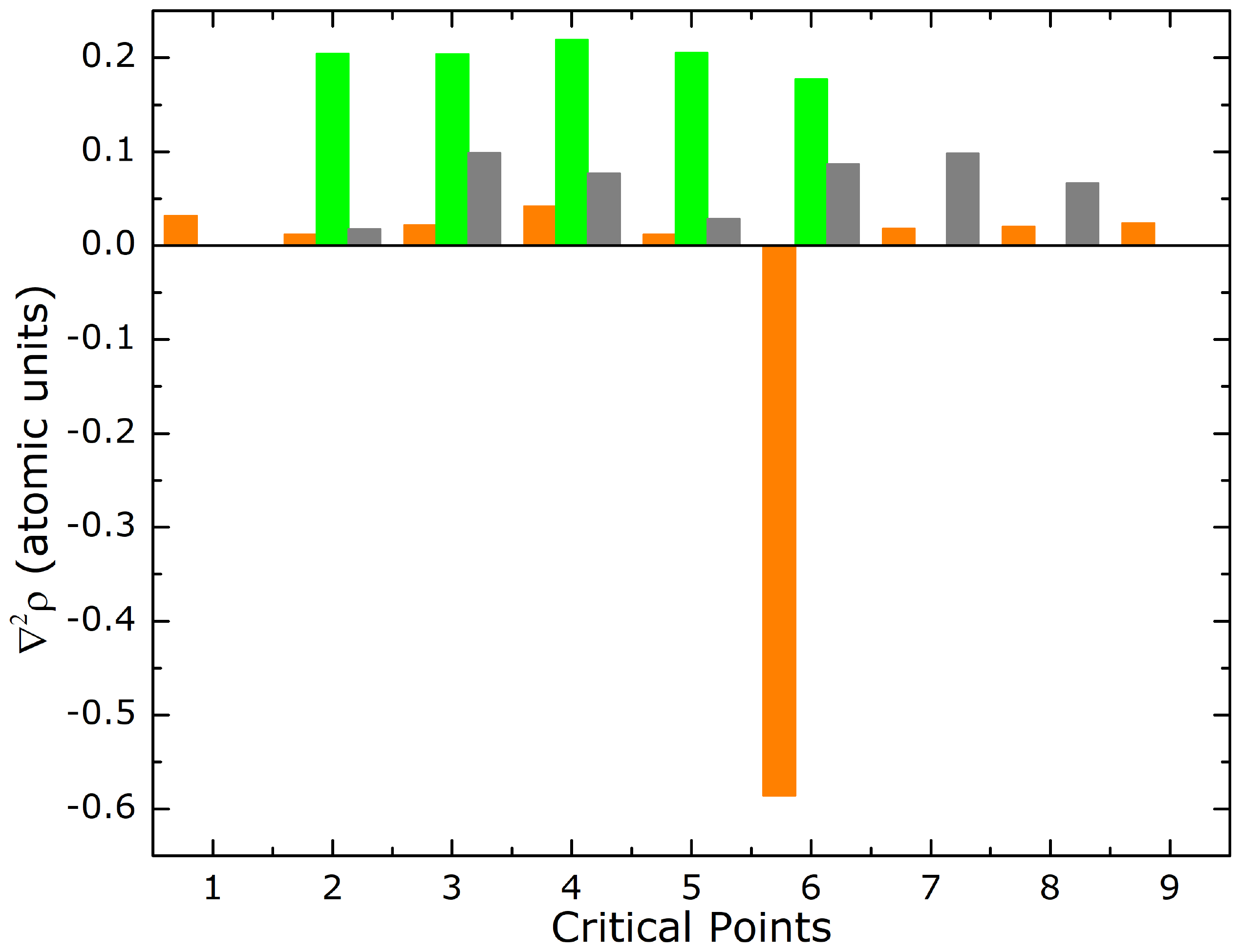}
\label{Fig:Lap_MCNB}}      &
\subfigure[MCNB+M4 (ELF)]{\includegraphics[width=\sizeA]{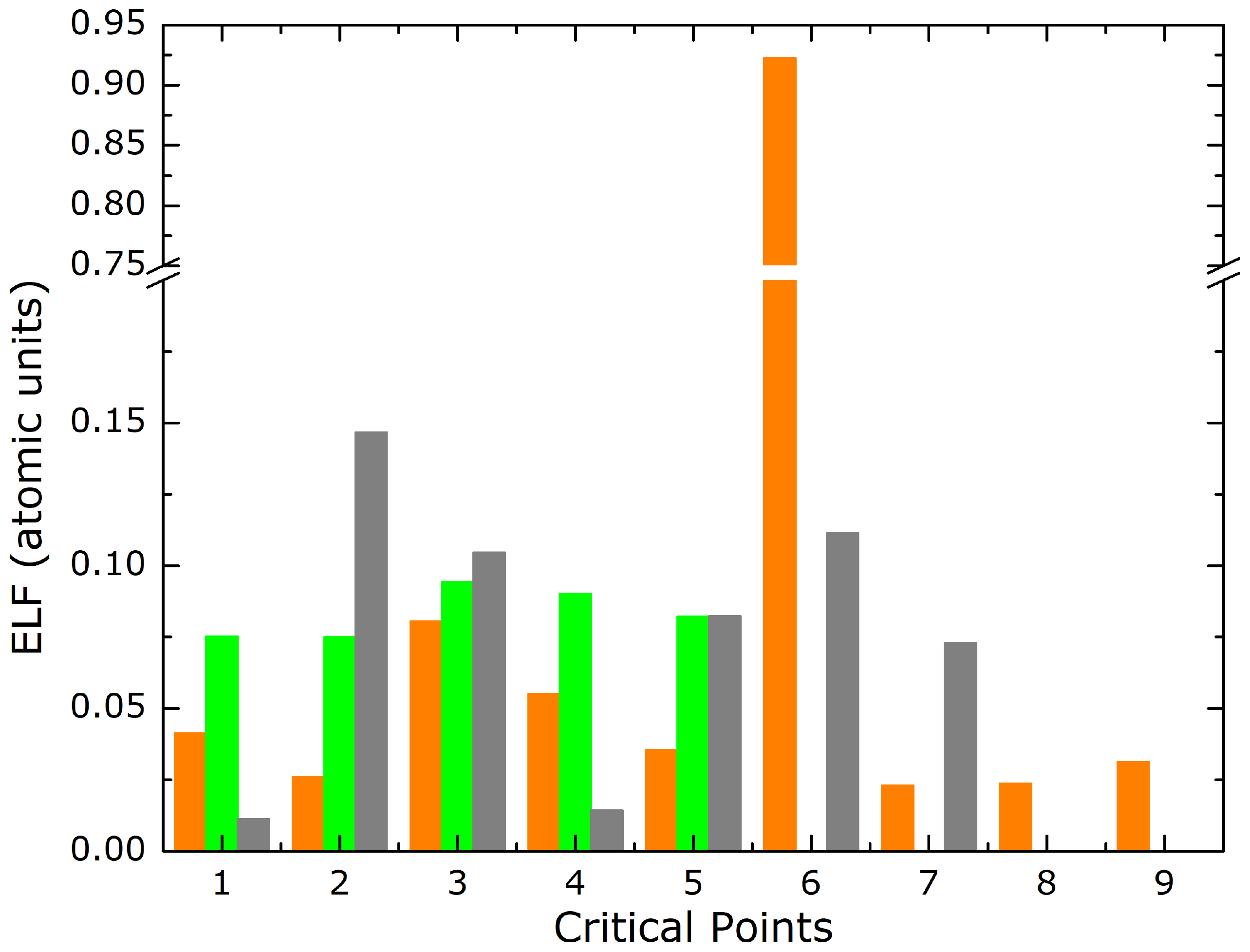}
\label{Fig:ELF_MCNB}}       &
\subfigure[MCNB+M4 (LOL)]{\includegraphics[width=\sizeA]{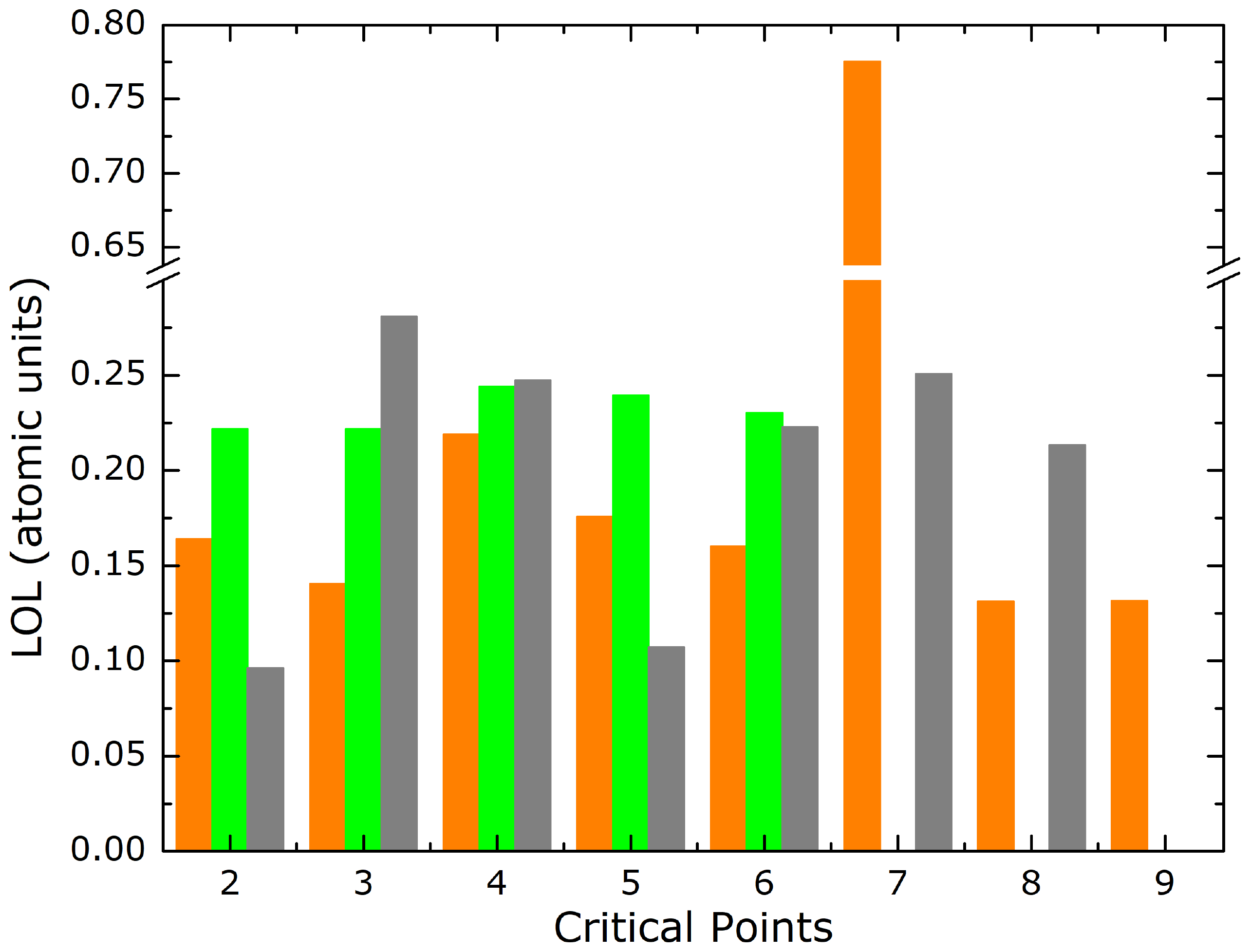}
\label{Fig:LOL_MCNB}} \\
\end{tabular}
\caption{\label{Fig:Topol} Topology results for Ni, Cd and Pb
nanoclusters are shown in green, orange and gray, respectively.}
\end{figure}

\section{Conclusions}
\label{Sec:Conclusions}
This study investigated the interactions between Ni, Cd, and Pb nanoclusters
and carbon nanobelts with different geometries. Various methods were used,
including the automated Interaction Site Screening (aISS) to identify the best
interaction regions, geometry optimization, molecular dynamics simulations,
electronic property calculations, and topological analysis.

The optimized structures were used to calculate the binding energy for each
complex, and the results indicated that the Ni nanocluster showed the most
favorable interaction followed by Pb and Cd nanoclusters. Molecular dynamics
simulations showed that the heavy metals remained bound to the nanobelts with
negative binding energy during the production time of 100~ps.

The electronic calculations revealed that the topology of the MCNB slightly
altered the HOMO/LUMO distribution, indicating good electron mobility. The
HOMO/LUMO surfaces were redistributed around the region where the metal was
located due to the metal/nanobelt interaction. Moreover, the Ni nanocluster
caused a more significant modification of the nanocluster's charges than the
other metals.

The topological analysis identified the critical points that helped
characterize the type and strength of the interactions. The MCNB had higher
values for all descriptors than the CNB systems, indicating that using MCNB to
adsorb heavy metals is a better choice. The Ni nanocluster was better adsorbed
than the Cd and Pb nanoclusters, as indicated by the values of the descriptors
used (electron density, Laplacian of the electron density, ELF, and LOL
indexes).

Overall, combining the results from geometry optimization, binding energy
calculation, and topological analysis, the study concluded that Ni nanoclusters
are chemisorbed, while Cd and Pb nanoclusters are physisorbed in both
nanobelts, with more favorable adsorption for the M\"obius carbon nanobelts.

\section*{CRediT authorship contribution statement}

\textbf{C. Aguiar}: Investigation, Formal analysis, Writing--original draft,
Writing--review \& editing. \textbf{N. Dattani}: Investigation, Resources,
Formal
analysis, Writing--original draft, Writing--review \& editing. \textbf{I.
Camps}:
Conceptualization, Methodology, Software, Formal analysis, Resources,
Writing--review \& editing, Supervision, Project administration.

\section*{Declaration of competing interest}

The authors declare that they have no known competing financial interests or
personal relationships that could have appeared to influence the work reported
in this paper.

\section*{Data availability}
The raw data required to reproduce these findings are available to download
from
\href{https://doi.org/10.5281/zenodo.7823747}{https://doi.org/10.5281/zenodo.7823747}.

\section*{Acknowledgements}
We would like to acknowledge financial support from the Brazilian agencies
CNPq, CAPES and FAPEMIG. Part of the results presented here were developed
with the help of a CENAPAD-SP (Centro Nacional de Processamento de Alto
Desempenho em S\~ao Paulo) grant UNICAMP/FINEP--MCT, CENAPAD--UFC (Centro
Nacional de Processamento de Alto Desempenho, at Universidade Federal do
Cear\'a), and Digital Research Alliance of
Canada (via  project bmh-491-09 belonging to Dr. Nike Dattani), for the
computational support.

\newpage

\end{document}